\documentstyle[12pt,aasms4]{article}
\begin{document}
\received{ }
\accepted{  }
\journalid{ }{ }
\slugcomment{   }

\lefthead{Barrado y Navascue\'es et al. }
\righthead{The Age of Beta Pic}

\title{The Age of Beta Pic\footnote{Based
on  observations collected by the Hipparcos satellite}}

\author{David Barrado y Navascu\'es,
\affil{Max-Planck Institut f\"ur Astronomie. K\"onigstuhl 17,
Heidelberg, D-69117, Germany}
\and 
 John R. Stauffer,
\affil{Harvard--Smithsonian Center for Astrophysics,
       60 Garden St., Cambridge, MA 02138, USA}
\and
Inseok Song, Jean-Pierre Caillault
\affil{Department of Physics and Astronomy, University of Georgia, Athens, 
GA 30602-24551, USA}
}


\begin{abstract} 

We have reanalyzed data for the proposed moving group associated
with $\beta$ Pictoris in order to determine if the group (or part of
it) is real, and, if so, to derive an improved age estimate
for $\beta$ Pic.   By using new, more accurate proper motions from
PPM and Hipparcos and a few new radial velocities, we conclude that on
kinematic grounds most of the proposed members of the moving group
are not, in fact, associated with $\beta$ Pic.   However, two M
dwarfs - or three, actually, since one of them is a nearly equal
mass binary - have space motions that coincide with that of $\beta$
Pic to within 1 km/s with small error bars.   Based on a color-magnitude
diagram derived from accurate photometry and Hipparcos parallaxes,
these two possible proper-motion companions to $\beta$ Pic are
very young; we derive an age of $\sim$20 Myr by comparison to theoretical
tracks from D'Antona \& Mazzitelli.  In fact, the proposed $\beta$ Pic
companions comprise two of the three youngest M dwarfs in the
sample of 160 dM stars for which we have data.   The chromospheric
and coronal activity of these two stars also confirm that they are
quite young.   We argue that the probability that two of the three
youngest nearby M dwarfs would accurately share the space motion
of $\beta$  Pic by chance is quite small, and therefore we believe that
$\beta$ Pic and the two M dwarfs (GL~799 and GL~803) were formed together.
The estimated 
 age for $\beta$ Pic is then 20$\pm$10 Myr, where the uncertainty
in the age arises primarily from possible errors in the PMS isochrones and in
the conversion from color to effective temperature.  
This young age for $\beta$ Pic  supports the contention that the 
IR-excess for the Vega-like stars is age  dependent.
\end{abstract} 
 

\section{Introduction}

This is the fourth paper in a series devoted to 
 the ages of Vega-like stars. This is 
achieved by finding late  type stars physically associated to
them. Then, several time-dependent properties are analyzed
and an age is derived. Stauffer et al. (\cite{Stauffer1995})
 studied the secondary  of HR4796, a
conspicuous Vega-like star discovered by Jura (\cite{Jura1991}).
They concluded that the binary is remarkably young
(8$\pm$2 Myr). More recently, Barrado y Navascu\'es et al. (\cite{ByN1997})
 focused on  Fomalhaut.
The physical association with the late  type star 
 GL~879 served to constrain the age to
200$\pm$100 Myr. The realization of the fact that Fomalhaut shares its
Galactic movement with other stars, including Castor and
 Vega, produced another
determination of the age,  
200$\pm$100 Myr (Barrado y Navascu\'es \cite{ByN1998}).

Vega-like stars show large
IR excesses originated in dusty circumstellar disks, which are
thought to be the remnants of the T Tauri disks or a consequence
of the formation of planets (see  Backman \& Paresce 
\cite{Backman1993}). Until the discovery of the first
extra-solar planet by Mayor \& Queloz (\cite{Mayor1995}), they provided
some of the best evidence for the presence of planetary systems outside 
of  our own.
There has been much recent progress on these systems in terms of understanding
the structure of their disks 
(Jura et al. 1998; Jayawardhana et al. 1998; Koerner et al. 1998; 
Greaves et al. 1998; Schneider et al. 1999), spectral distribution
(Zuckerman \& Becklin 1993; Holland et al. 1998;
Fajardo-Acosta et al. 1998) 
and evolution (Zuckerman et al. 1995; Thakur et al. 1997; Song et al. 1998).
 However, it is still true that
accurate ages for these systems are still in considerable short supply.
In this paper, we provide what we believe to be an accurate age for
 $\beta$ Pic.

\section{A common origin based on the kinematic properties}

Following  Barrado y Navascu\'es (\cite{ByN1998}),
 we  selected an initial list of possible 
 $\beta$ Pic companions from Agekyan \& Orlov 
(\cite{Agekyan1984}), which provides a extensive search of kinematic groups. 
We also included stars  from Soderblom (\cite{Soderblom1990}),
Poveda et al. (\cite{Poveda1994}) and Tokovinin (\cite{Tokovinin1997}).
Then,  we computed the  Galactic components of these stars using equatorial
 coordinates, parallaxes, proper motions 
--the Hipparcos
(ESA, \cite{ESA1997}) and PPM 
(Bastian \& Roeser \cite{Bastian1993}; Roeser \&  Bastian 
\cite{Roeser1994}) catalogs--
and the radial velocities (Duflot et al. \cite{Duflot1995}).
For $\beta$ Pic itself, we used the values derived by 
Lagrange et al. (1995)  and Jolly et al. (1998),
based on HST/GHRS spectra of narrow absorption lines of Fe and
CO (thought to be due to stationary circumstellar gas 
but external to its disk).
Of the initial stars selected, only six have space motions
within 2 sigma  of that of  $\beta$ Pic to be
 at all plausible companions. 
Tables 1 and 2 provide various data for these stars and  the dynamics.
The UVW components of the Galactic velocity  were computed following
Johnson \& Soderblom (1987), using PPM proper motions.
 Similar results can be computed with Hipparcos.
We have used these data for two different
purposes: First, we have tried to verify if indeed these stars are physically
associated. Second, using several properties of the late  type
stars, we have estimated the age of the moving group.

The V component 
imposes the strongest constraint
to determine whether a group of stars are associated
 (Soderblom \& Mayor \cite{Soderblom1993}). 
V  should correspond to a drift
rate which would lead to a secular increase in separation between two
given stars (as opposed to U or W components, where
 a difference in current velocity may not matter, because stars  oscillate
in those directions).  Since  2 km/s is about 2 pc/Myr, 
two stars whose space motion differed by that amount would
separate by 40 pc in 20 Myr
 (our final estimation of the age of $\beta$ Pic).
 Therefore,  they could not both have been born 
in the same place 20 Myr ago and now both be within 20 pc of the Sun.
Given the accuracies to which we can estimate the space
motions of our selected stars, we believe that GL~97 can be
rejected as a possible companion to $\beta$ Pic.  If we use
the PPM proper motions, it has too large a difference
in the V component; if instead, we
adopt the Hipparcos proper motions, then the U velocity
differs by an amount ($>$8 km/s) that is too large.  There
are other spectroscopic reasons to also believe that
GL~97 is too old to be a possible companion to $\beta$ Pic
(Pasquini et al. \cite{Pasquini1994}).
We also choose to exclude GL 601 from further consideration
because we have no observational data that allows us to
usefully estimate its age.  In the next section, we will 
examine age estimates for Beta Pic and for the remaining
four stars in order to try to establish whether any of
them appear to be coeval.

\section{The age of $\beta$ Pic}

\subsection{Isochrone Fitting for $\beta$ Pic Itself}

Isochrone fitting for $\beta$ Pic  has been  previously attempted,
 yielding an age of 100 Myr 
(Backman \& Paresce \cite{Backman1993}). Lanz et al. (\cite{Lanz1995}), using 
the same procedure, concluded that an age around 12 Myr 
 or larger than 300 Myr would be possible,
 although they judged the first  value as the  most likely. 
From Figure 2 of Brunini \& Benvenuto (\cite{Brunini1996}),
which represents evolutionary tracks,
an age between 20 and 40 Myr could be inferred.
Finally, Crifo et al. (\cite{Crifo1997}),
 using Hipparcos data, confirmed that the star is 
very close to or on the ZAMS,
 and it is older than 8 Myr.
All these studies show that this technique is not very restrictive
and the age of $\beta$ Pic remains uncertain.

\subsection{Isochrone Fitting for the Possible Companions of $\beta$ Pic}

The photometry
of late type stars can provide accurate ages, if they are cool and young 
enough to be in the PMS phase, by comparison with isochrones.
We have compared the four candidate $\beta$ Pic companions
 to PMS isochrones from  D'Antona \&
Mazzitelli (1997, priv. comm.=DM97), 
where we have used a color-temperature conversion
based on requiring the DM97 125 Myr isochrone to coincide with the
main-sequence locus of stars in the Pleiades (c.f. Stauffer et al. 
\cite{Stauffer1995}; Stauffer \cite{Stauffer1998}).
In order to place our four candidate $\beta$ Pic companions in context,
we also include in this figure all of the nearby M dwarfs for which
Leggett (1992) has compiled accurate photometry and which have parallaxes
from Hipparcos with $\sigma(\pi)$/$\pi$ $<$ 0.10.  For binary stars
where it is known that the two components are nearly equal in brightness
(including GL~799), we have added 0.75 mag to the M$_V$\ in order to
correct the binarity effect;
for know binaries with unknown mass ratios,  we plot the
star as an open diamond symbol but do not correct the M$_V$.
Clearly, GL~799 and GL~803 are among the brightest stars,
compared with stars of the same color, 
in the solar neighborhood. That is, they are very young. 
In fact, the three youngest objects in Figure~\ref{fig1} are
GL~182, GL~799 and GL~803.  GL~182 is known to be a very young, dMe
star (Favata et al. \cite{Favata1998});
 its kinematics indicates that it is not,
however, moving with the same space motion of GL~799 and 803, so we
do not consider it further.  Within the errors, the locations of GL~799
and 803 in Figure~\ref{fig1} are consistent with their having the
same age (20 Myr).   GL~781.2 and GL~824 appear to
be older, with ages formally consistent with being 40 Myr.
  However, because they are
higher mass objects and their displacement above the ZAMS is less,
their locations in Figure~\ref{fig1} are actually consistent with
any age up to several hundred Myr given uncertainties in their
photometry and  metallicity and the placement of the isochrones
into the observational plane.  We provide evidence in the next section
that GL~824, at least, is quite old ($>$ 600 Myr), and 
unlikely to be physically connected to $\beta$ Pic.

\subsection{Stellar activity}

Stellar activity, a consequence of the presence of magnetic fields
in late--type stars (due to the combination of the rotation and convection,
or dynamo effect), is a well studied phenomenon.  Because of main
sequence angular momentum loss, the rotation rates of low mass stars
decline with age - and hence  activity levels
also decline with age (e.g., Stauffer \cite{Stauffer1988}). 
We can use measures of stellar activity, therefore, as proxies
for age in an attempt to identify which of our candidate stars might
be coeval with $\beta$ Pic.
Figure~\ref{fig2} depicts the X-ray luminosity
against  (B--V). In panel a, 
crosses represent ROSAT All Sky data (H\"unsch et al. \cite{Hunsch1999})
for the Gliese stars.
In panel b, Pleiades and Hyades stars appear as open and solid symbols, 
respectively. Clearly, the X-ray luminosity of 
GL~824 is relatively low, even compared to that of the stars in the Hyades
(age$\sim$600 Myr); we infer from this that GL~824 is older than the Hyades.
Based on its location in a CM diagram, $\beta$
Pic cannot be as old as 600 Myr, and therefore the X-ray data provide
strong evidence that GL~824 is older than $\beta$ Pic.
On the other hand, the  activity  of GL~803 and GL~799
is very high, consistent with the young ages deduced  from
their position in the CM diagram.   Unfortunately, there
is no published X-ray  for GL781.2, and
we cannot further constrain its age at this time.

\subsection{Summary of Age Constraints}

Using the best available data, analysis of the location of $\beta$ Pic
in a CM diagram only allows one to conclude that its age
is somewhere between a few million  and a few hundred million years.
Of the four late  type stars
 whose kinematics match that of $\beta$ Pic, GL~824
is removed from contention because its  activity  indicates
it is  older than the maximum age for $\beta$ Pic.   
 GL~781.2  is essentially unconstrained in its age due to a lack
of activity data and due to its early spectral
type (precluding an accurate HR diagram age.  However,
the other two candidates have a well-constrained age from their location
in a CM diagram, have  activity
levels consistent with that age, and share the motion of $\beta$ Pic
to within 1 km/s.  We believe, therefore, that the age derived for
these stars from PMS isochrones - 20$\pm$10 Myr - is the best
estimate for the age of $\beta$ Pic.   The spatial location of the three
stars is compatible with this age and the derived relative space motions.
We note that Poveda et al. (1994) has previously identified GL~799 and
GL~803 as being likely siblings - we are now simply adding $\beta$
Pic as their bigger brother.

\section{The Correlation of IR Excess and Age for $\beta$ Pic Stars}

In their comprehensive review of the Vega phenomenon, Backman \&
Paresce (\cite{Backman1993}) described specifically the
evolutionary status of the three prototypes, $\beta$ Pic, Vega and
Fomalhaut, estimating their ages as 100, 200 and 400 Myr, respectively.
Several studies have tried to relate these ages with different 
properties which appear as a consequence of the presence of  circumstellar 
disks, in order to see if there is an evolutionary sequence.
For instance, Fig. 2 of Holland et al. (1998) suggests a dependence with 
the age of the total amount of dust in the disk. 
Our results support this type of dependence (Figure~\ref{fig3}).  
The inferred rapid decline in dust mass supports
the hypothesis that the
Vega phenomenon is a normal stage in the early life of intermediate mass and
solar-like stars.

\acknowledgements

 DBN thanks the  IAC (Spain) and the
DFG (Germany) for  
their fellowship. JRS acknowledges support from NASA Grant
NAGW-2698 and 3690.
We  thank X. Delfosse and D. Fischer for providing data  prior to publication,
We have used the Simbad database.


\small{
\begin{table}
\caption[ ]{Photometry and other data for the  $\beta$ Pic Moving Group.}
\begin{tabular}{lcccccccl}
\hline
 Gliese  & Sp. Type & V    &  Mv   & (B-V)  & (V-I)c  &Log Lx$^1$&EW(H$_\alpha$)$^2$ & Other names \\
  number &          &      &       &        &         & (erg/s) &  (\AA)           &                    \\
\hline
\hline
219      & A3V      & 3.85 & 2.425 & 0.171  & 0.18    & --      & --       & HD39060,  $\beta$ Pic    \\
\hline
97       & G2V      & 5.19 & 3.485 & 0.608  & 0.68    & 29.60   & --       & HD14802,  $\kappa$ For   \\
601 A    & F2III    & 2.83 & 2.38  & 0.30   & 0.36    & --      &  --      & HD141891, $\beta$ TrA    \\
781.2    & K3/K4V   & 9.75 & 7.241 & 1.137  & 1.249   & --      & --       & HD191285,                \\
799AB &M4.5e+M4.5e&10.27&10.97$^3$&1.550$^3$ &2.90$^3$ &29.55$^3$& 4.10    & HD196982,  AT Mic         \\
 803     & M1Ve     & 8.81 & 8.82  & 1.470  & 2.10    & 29.74   &  1.56    & HD197481,  AU Mic         \\
 824     & K2       & 7.88 & 6.84  & 1.020  & 1.11    & 27.76   & -0.91    & HD202575,                \\
\hline 
\end{tabular}
$\,$    \\
Sp. Type and photometry from Hypparcos and Bessel (1990),
$^1$ Lx values from H\"unsch et al. (1999),
$^2$ EW(H$\alpha$) from Panagi \& Mathioudakis (1993),
$^3$ Values for each individual component.
\end{table}
}

\footnotesize{
\begin{flushleft}
\begin{table*}
\tiny
\caption[ ]{Coordinates and velocities for the $\beta$ Pic Moving Group.}
\begin{tabular}{ccccccrrrrr}
\hline
GL  &\multicolumn{1}{c}{alpha}&\multicolumn{1}{c}{delta}&\multicolumn{1}{c}{parallax}
&\multicolumn{1}{c}{$\mu_{\alpha}$}  & \multicolumn{1}{c}{$\mu_{\delta}$} &\,&
\multicolumn{1}{c}{RV}& \multicolumn{1}{c}{U} & \multicolumn{1}{c}{V} & \multicolumn{1}{c}{W} \\
\cline{2-3} \cline{5-6} \cline{8-11}
 &\multicolumn{2}{c}{(h m s)(1950.0)($^\circ$ ' '')}&\multicolumn{1}{c}{(mas)}
 &\multicolumn{2}{c}{(mas/yr)}&& \multicolumn{4}{c}{(km/s)}\\
\hline
\hline
219   &  5:46:05.93 & -51:05:01.8 & 51.87$\pm$0.51 &   9.4$\pm$4.2   &  79$\pm$4.3     && 20.0$\pm$0.5$^3$     &-10.6$\pm$0.4 &-16.3$\pm$0.5 & -8.7$\pm$0.4\\
 ``   &       ``    &      ``     &      ``        &        ``       &      ``         && 21.0$\pm$1.0$^4$     &-10.7$\pm$0.4 &-17.1$\pm$0.9 & -9.2$\pm$0.6\\
\hline
 97   &  2:20:15.23 & -24: 2:36.2 & 45.60$\pm$0.82 &197.34$\pm$0.77$^*$&-4.39$\pm$0.51$^*$&&18.6$\pm$1.0$^2$   &-19.3$\pm$0.4 &-16.9$\pm$0.3 &-10.4$\pm$0.9\\
 ``   &       ``    &      ``     &       ``       & 200.0$\pm$0.6   & -62$\pm$0.8     &&     ``               &-15.5$\pm$0.4 &-21.5$\pm$0.3 &-10.8$\pm$0.9\\
601   & 15:50:42.96 &-63:16:42.7  &81.24$\pm$0.62  & -192.2$\pm$0.9  & -398$\pm$1.0    &&  0.4$\pm$1.0$^2$     &-15.3$\pm$0.8 & -17.8$\pm$0.6 & -10.7$\pm$0.2\\
781.2 & 20:06:47.92 & -14:26:00.2 & 31.49$\pm$1.58 &  81.3$\pm$2.1   & -85$\pm$2.1     &&-12.3$\pm$3.0$^2$     &-12.5$\pm$2.5 &-14.4$\pm$1.5 &-10.1$\pm$1.4\\
799AB & 20:38:43.71 & -32:36:33.8 & 97.80$\pm$4.65 & 274.2$\pm$2.9   &-351$\pm$3.0     && -3.5$\pm$1.0$^2$     & -9.7$\pm$0.9 &-16.1$\pm$0.8 &-11.2$\pm$0.8\\
803   & 20:42:03.79 & -31:31:05.6 &100.59$\pm$1.35 & 281.3$\pm$2.9   &-363$\pm$2.9     &&-4.89$\pm$0.02$^1$    &-10.5$\pm$0.2 &-16.6$\pm$0.3 &-10.4$\pm$0.2\\
824   & 21:14:05.39 &  09:11:09.3 & 61.83$\pm$1.06 & 143.6$\pm$2.2   &-121$\pm$2.1     &&-13.2$\pm$1.0$^2$     & -8.3$\pm$0.5 &-16.2$\pm$0.8 & -7.0$\pm$0.5\\
\hline
\end{tabular}
$\,$\\
$^*$ Hipparcos, all others from PPM\\
RV from: $^1$ Delfosse (1999, priv.comm.), $^2$ Duflot et a. (1995) 
$^3$ Lagrange et al. (1995), $^4$ Jolly et al. (1998).\\
\end{table*}
\end{flushleft}
}

\newpage

\normalsize

\begin{figure*}
\vspace{1cm}
\caption{Color-Magnitude diagram for the late spectral type candidates.
The isochrones (20, 30, 50 and ZAMS) 
are those from D'Antona \& Mazzitelli (1997). Stars from Leggett (1992)
are shown as crosses. Open diamonds represent the combined photometry
of binaries.}
\label{fig1}
\end{figure*}

\begin{figure*}
\vspace{1cm}
\caption{X-ray luminosities plotted versus  the (B-V) color indices.
{\bf a} Gliese stars.
{\bf b} Pleiades and Hyades members. Triangles indicate upper limits.
}
\label{fig2}
\end{figure*}

\begin{figure*}
\vspace{1cm}
\caption{Fractional infrared luminosity versus age.}
\label{fig3}
\end{figure*}

\newpage

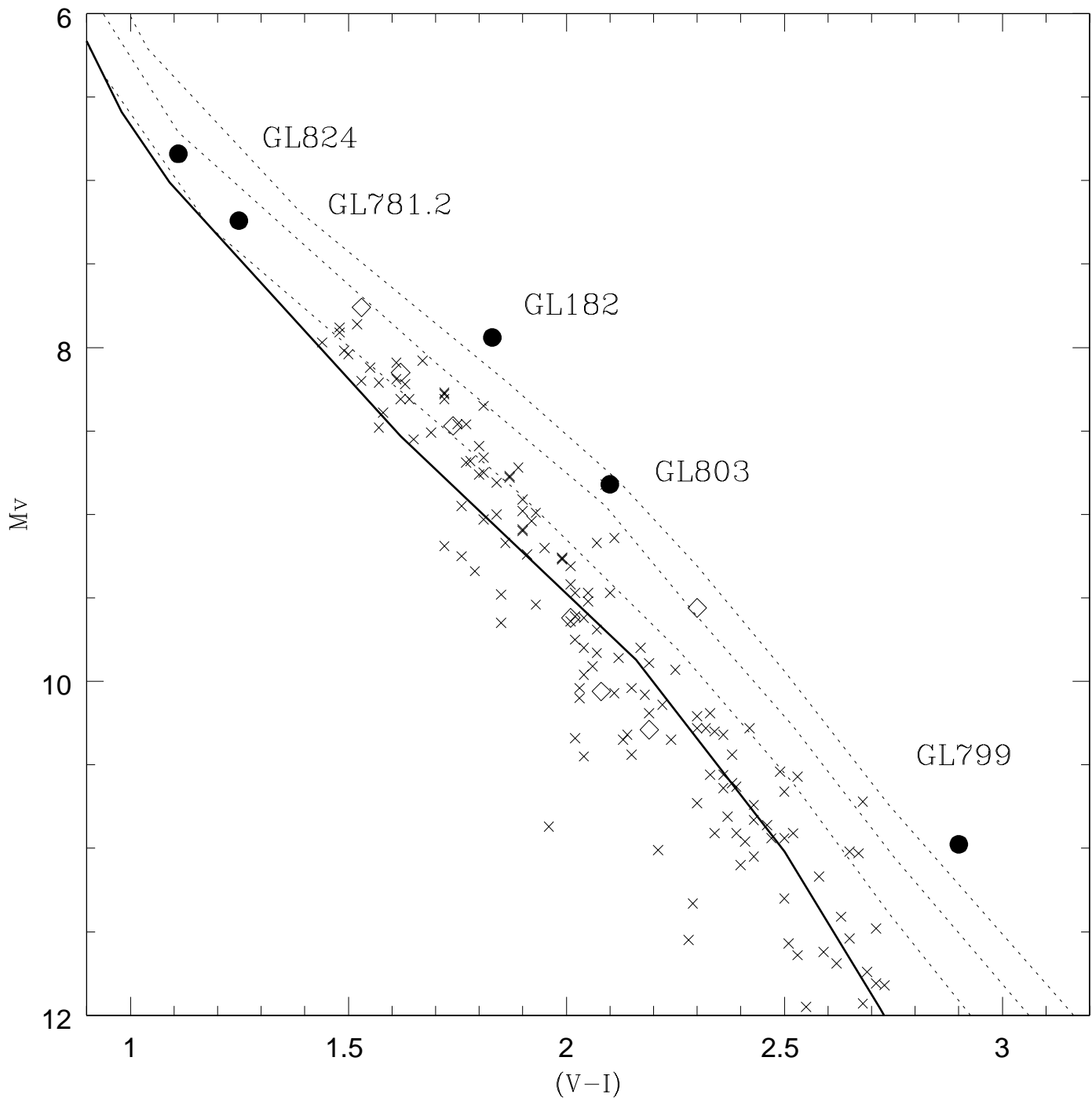
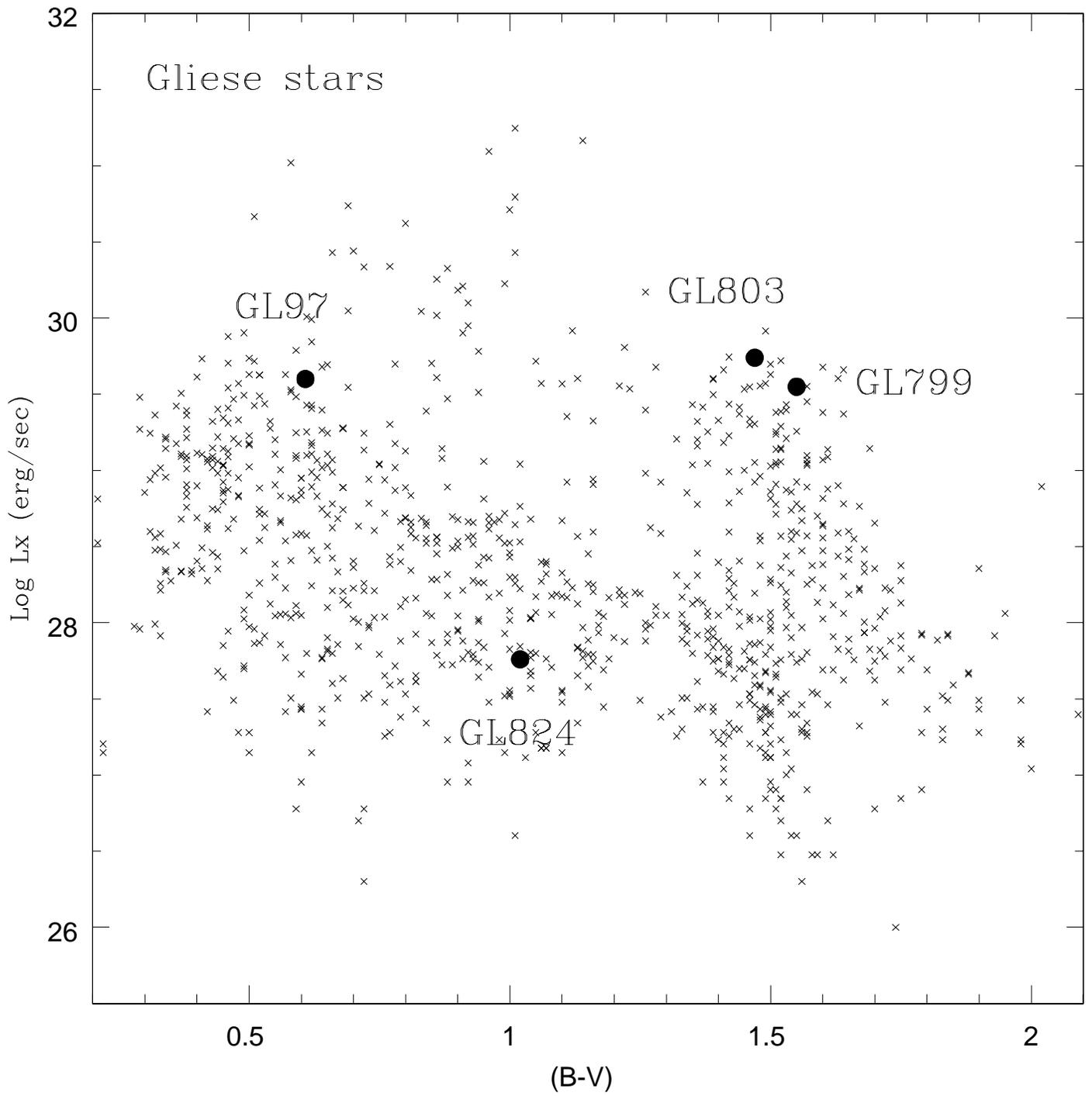
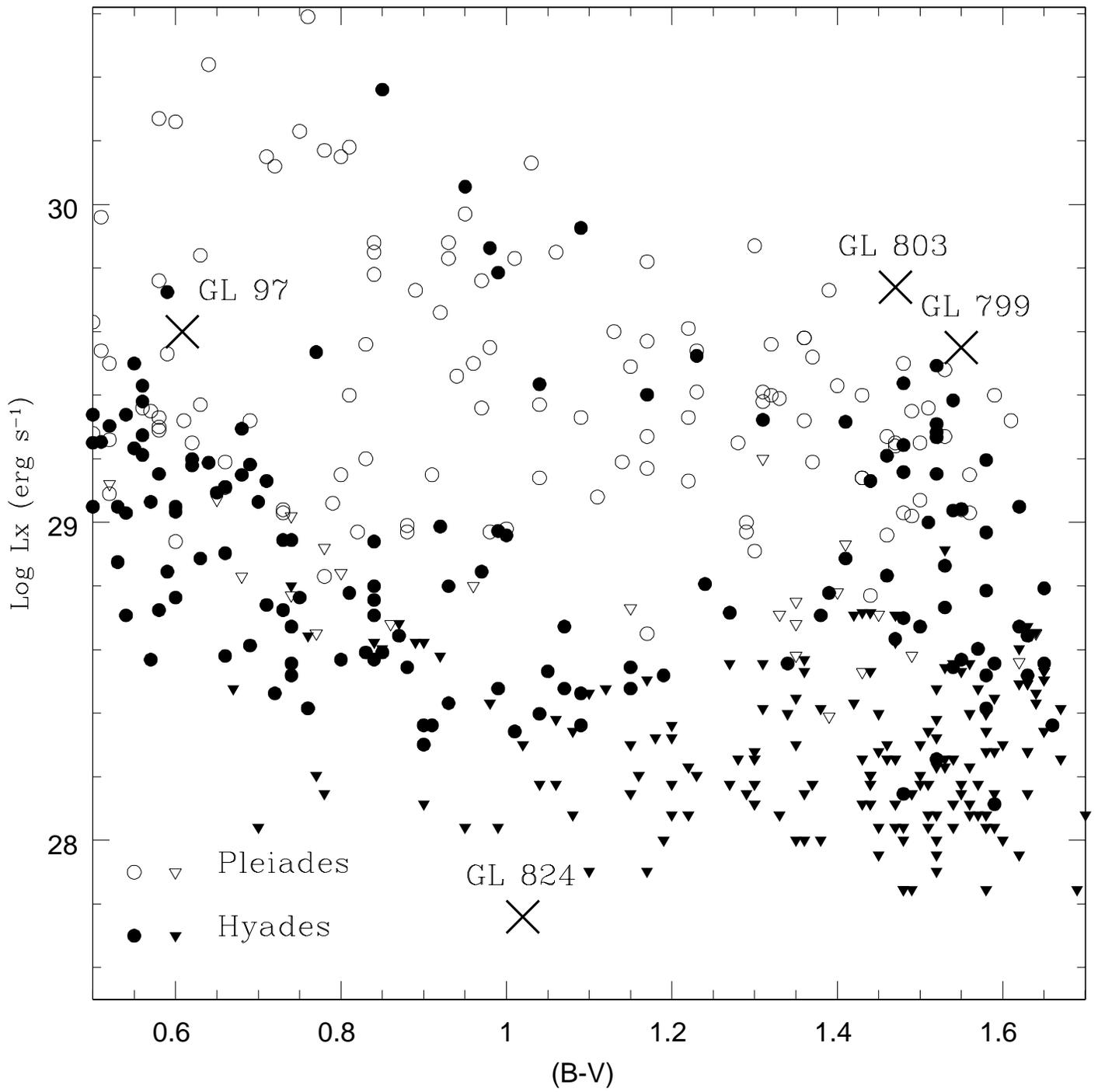
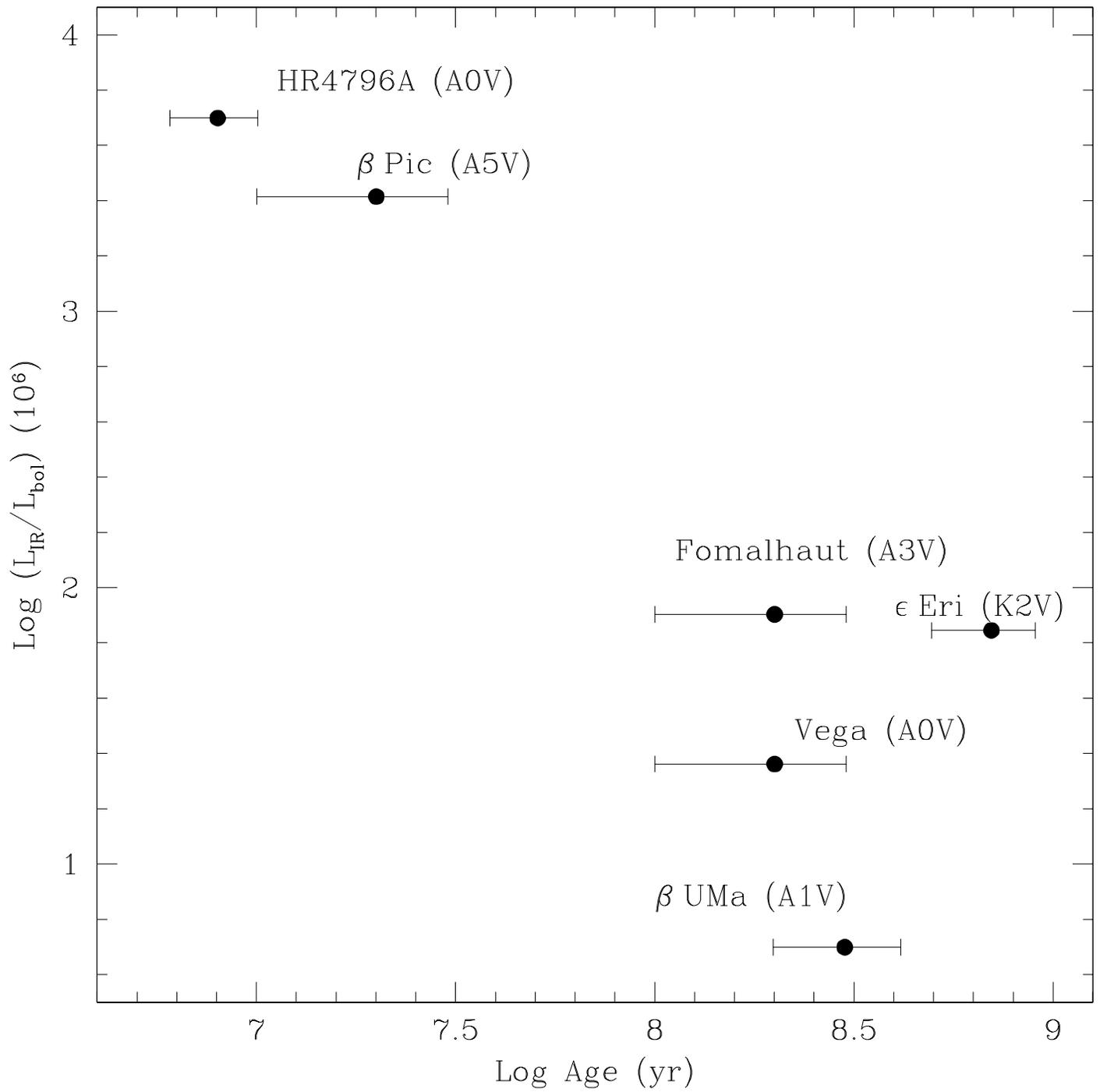
\end{document}